\documentclass[pre,twocolumn,twoside,showpacs,byrevtex,superscriptaddress]{revtex4}

\usepackage{array}

\newcommand{\PreserveBackslash}[1]{\let\temp=\\#1\let\\=\temp}
\let\PBS = \PreserveBackslash

\usepackage{graphicx,epsfig,verbatim,enumerate}
\usepackage{amssymb,amsmath}
\usepackage{ifthen}
\newboolean{twocolswitch}

\usepackage{color}

\newcommand{\sindex}[1]{}
\newcommand{\nindex}[1]{}

\newcommand{\www}[1]{\url{#1}}

\newcommand{\Req}[1]{Eq.~(\ref{#1})}

\newcommand{\veck}{\vec{k}}

\newcommand{\Prob}{\mathbf{Pr}}
\newcommand{\Probof}[1]{\mathbf{Pr}(#1)}

\newcommand{\bidmark}{\rm u}
\newcommand{\inmark}{\rm i}
\newcommand{\outmark}{\rm o}

\newcommand{\kin}{k_{\inmark}}
\newcommand{\kout}{k_{\outmark}}
\newcommand{\kbid}{k_{\bidmark}}
\newcommand{\keff}{k^{(\rm eff)}}

\newcommand{\Probin}{P^{(\rm \inmark)}}
\newcommand{\Probout}{P^{(\rm \outmark)}}
\newcommand{\Probbid}{P^{(\rm \bidmark)}}

\newcommand{\probfc}{Q}

\newcommand{\nodetype}{\nu}
\newcommand{\edgetype}{\lambda}

\newcommand{\supone}{{(1)}}
\newcommand{\suptwo}{{(2)}}
\newcommand{\setone}{A^{\supone}}
\newcommand{\settwo}{A^{\suptwo}}

\setboolean{twocolswitch}{true}

\begin{document}

\title{
  Direct computation of contagion triggering probabilities for generalized and bipartite random networks

}

\author{
\firstname{Kameron Decker}
\surname{Harris}
}

\email{kamdh@uw.edu
}

\affiliation{
  Applied Mathematics,
  University of Washington,
  Lewis Hall \#202, 
  Box 353925, 
  Seattle, WA 98195-3925.
}

\author{
\firstname{Joshua L.}
\surname{Payne}
}

\email{joshua.payne@ieu.uzh.ch}

\affiliation{
Institute of Evolutionary Biology and Environmental Sciences,
University of Zurich,
Winterhurerstrasse 190,
8057, Zurich, Switzerland.
}

\author{
\firstname{Peter Sheridan}
\surname{Dodds}
}

\email{peter.dodds@uvm.edu}

\affiliation{
  The University of Vermont,
  Burlington, VT 05401.}

\affiliation{Complex Systems Center,
  Computational Story Lab,
  the Vermont Advanced Computing Core,
  \& Department of Mathematics \& Statistics,
  The University of Vermont,
  Burlington, VT 05401.}

\markboth{Title}
{Author names}

\date{\today}

\begin{abstract}
  We derive a general expression for the probability of global
spreading starting from a single infected seed for contagion processes
acting on generalized, correlated random networks.
We employ a simple probabilistic argument that 
encodes the spreading mechanism in an intuitive, physical fashion.
We use our approach to directly and systematically 
obtain triggering probabilities for contagion
processes acting on a collection of random network families
including bipartite random networks.
We find the contagion condition,
the location of the phase transition into an endemic state,
from an expansion about the disease-free state.  
\end{abstract}


\maketitle

\section{Introduction}
\label{sec:gcgrn.introduction}

Spreading is a pervasive dynamic phenomenon,
ranging in form from simple physical diffusion to 
the complexities of socio-cultural
dispersion and interaction of ideas and beliefs~\cite{richerson2005a,chmiel2011a,romero2011a,rozin2001a,leskovec2009a,berger2009a,banerjee1992a,barsade2002a,bikhchandani1992a,rogers2003a,sieczka2011a}.
Successful spreading in systems may manifest
as an expanding front, such as in the spread
of disease through medieval Europe~\cite{cliff1981a},
or through inherent or revealed networks,
such as in pandemics in the modern era of global
travel~\cite{colizza2007a}. 
Here, we focus on spreading processes
operating on generalized random networks,
which have proven over the last decade
to be 
illustrative of spreading on real networks
and at the same time to be
analytically tractable~\cite{newman2001b,boguna2005a,meyers2006a,gleeson2007a,gleeson2008a,gleeson2010a,watts2002a,hackett2011a,ikeda2010a,romero2011a,munz2009a,watts2009a}.

In contributing to the wealth
of already known results for 
contagion on random networks,
we make two main advances here.
First, we obtain, in the most general terms possible,
an expression for the probability
of global spreading 
from a single seed for a broad range of 
contagion processes acting on generalized, correlated
random networks.  
By global spreading we mean
a non-zero fraction of nodes
in an infinite network are eventually infected.
Second, we use an argument
that is physically motivated and direct.
Existing approaches rely on a range
of mathematical techniques,
such as probability generating functions~\cite{wilf2006a,newman2001b,newman2003a},
which, while being entirely successful in
determining spreading probabilities and higher moments
of cascade size distribution,
obscure the underlying physical mechanisms.

\begin{figure}[htbp!]
  \centering
  \includegraphics[width=\columnwidth]{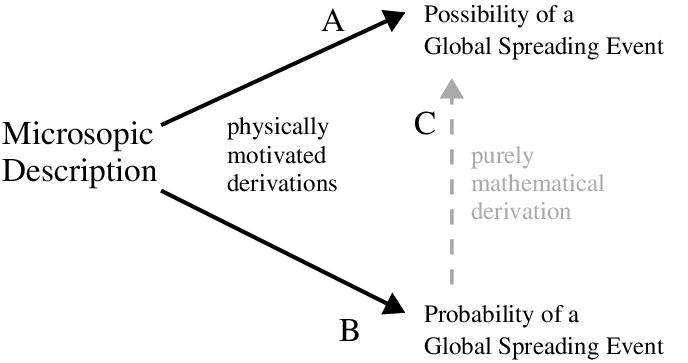}
  \caption{
    Physical and mathematical explanations
    of two fundamental aspects of broad classes of contagion processes
    acting on generalized random networks.
    In the present paper, we provide a physical approach
    to determining the probability of spreading from
    a single seed (derivation B).
    We use mathematical arguments to arrive again at 
    the binary contagion condition (derivation C),
    which we obtained in a previous work~\cite{dodds2011b}
    using a direct physical explanation (derivation A).
  }
  \label{fig:gcgrn.connection}
\end{figure}

The present paper is a companion
to our earlier work where 
we derived a general condition for
the possibility (rather than probability) 
of global spreading for single-seed contagion
processes acting on random networks~\cite{dodds2011b}.
We used specific results from both works in a
separate investigation of exactly solvable
network spreading models~\cite{payne2011a}.
As we show below, our expression
for the probability of spreading naturally
allows us to recover our expression for the possibility of
spreading, and this is a purely mathematical exercise.
Our key contribution is the direct derivation
of triggering probabilities via physical arguments,
as illustrated in Fig.~\ref{fig:gcgrn.connection}.

We structure our paper as follows.
In Sec.~\ref{sec:gcgrn.generalizedrandomnetworks},
we define the broadest class of
correlated random networks allowing
for directed and undirected edges and
arbitrary node and edge properties.
In Sec.~\ref{sec:gcgrn.contagion},
we define the general class of
contagion processes that our
treatment can encompass.
In Sec.~\ref{sec:gcgrn.analysis},
we compute the probability that seeding a node
of a given type generates a global spreading 
event.
For completeness, in Sec.~\ref{sec:gcgrn.contagioncondition},
we derive the contagion condition 
(location of the endemic phase transition)
result found in~\cite{dodds2011b},
and we show how non-physical expressions may 
arise through this mathematical route.
We use our formalism for six interrelated random network families
with general contagion processes acting on them in
Sec.~\ref{subsec:gcgrn.applicationrandom}.
In Secs.~\ref{subsec:gcgrn.applicationbipartite}
and~\ref{subsec:gcgrn.simplebipartite},
we show how our approach readily applies
to random bipartite networks,
and we offer some concluding remarks in
Sec.~\ref{sec:gcgrn.conclusion}.

\section{Generalized random networks}
\label{sec:gcgrn.generalizedrandomnetworks}

\begin{figure}[h!]
  \centering
  \includegraphics[width=0.49\textwidth]{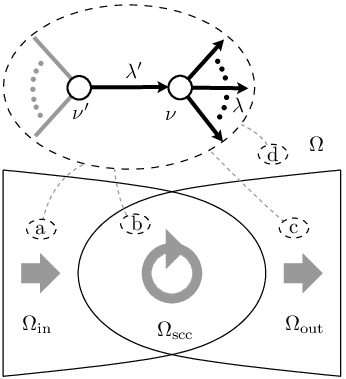}
  \caption{
    Schematic showing the configuration of the potential triggering node
    subnetwork
    using the present work's formalism for generalized random networks
    described in Sec.~\ref{sec:gcgrn.generalizedrandomnetworks}, and the basic
    form of a random network with directed edges and a giant
    component.
    The ellipses labelled a--d show four possible locations
    of the subnetwork in the overall network $\Omega$.
    Global spreading events can be successfully generated
    only if the subnetwork is part of the giant in-component $\Omega_{\rm in}$,
    either within or outside of 
    the giant strongly connected component $\Omega_{\rm scc}$ (ellipses a and b).
    No spreading is possible if the subnetwork is
    instead part of the giant out-component outside of
    the strongly connected component ($\Omega_{\rm out} / \Omega_{\rm scc}$, ellipse c)
    or outside of all three giant components (ellipse d).
  }
  \label{fig:gcgrn.sketch}
\end{figure}

Our theoretical treatment builds
on a formalism we introduce here for representing
generalized random networks, an expansion
of what we used in our connected, earlier work~\cite{dodds2011b}.
Our theory applies to large random networks with bounded degrees
(such as the configuration model),
since these graphs are all locally tree-like and can be approximated
by multitype branching processes.
Generalized random networks may contain a combination
of directed and undirected edges, 
so they are in general nonsimple graphs.

We depict the essential features
of a random network with possibly directed edges
in Fig.~\ref{fig:gcgrn.sketch}, noting 
that our analytic treatment will also cover
more specialized random networks, such
as those induced by bipartite graphs,
or networks with multipartite structure
(see Sec.~\ref{subsec:gcgrn.applicationbipartite}).
The most basic elements of networks are nodes and edges, 
and here we allow the following features encoded in
two types of labels:
\begin{itemize}
\item
  Node type, $\nu \in \mathcal{N}$: 
  arbitrary node characteristics
  such as node age, susceptibility to 
  a given disease or message, etc.
  The node type implicitly includes information
  about its degree, which
  we explain below.
\item
  Edge type, $\lambda \in \Lambda$: 
  arbitrary edge characteristics
  such as age, strength, conductance, etc.
  Since edges may be directed, 
  edge type 
  includes whether an edge is directed or not and
  its orientation if so.
  We thus use the notation $\bar{\lambda}$
  to indicate the edge's type when considered
  as travelling in the disallowed direction.
  (There is no need to distinguish $\lambda$ or $\bar{\lambda}$
  for undirected edges.)
  In other words, if there is a directed edge of type
  $\lambda$ from node $u$ to node $v$,
  we say there its type is $\bar{\lambda}$
  when viewing $v$ as the source and $u$ as the target.
\end{itemize}
We take $\mathcal{N}$ and $\Lambda$ 
to be discrete.
We denote the entire network by $\Omega$,
and the set of edge types incident to a node of type $\nu$
by $\Lambda_{\nu}$.

We define degree as the number of edges 
of a certain type emanating from a node.
In simple networks, we let
$k(\nu, \lambda)$ 
denote the number of edges of type $\lambda$ 
emanating from a node of type $\nu$.
In more general networks
we let the multi-index
$\veck(\nu, \lambda) = 
\left[ \kbid(\nu, \lambda), \kin(\nu, \lambda), \kout(\nu, \lambda) \right]$
denote the number of 
undirected, inward, and outward edges of type $\lambda$
belonging to a node of type $\nu$.

The `total degree' of a node of type $\nu$ is then
$\veck(\nu) = \sum_{\lambda \in \Lambda_\nu} \veck(\nu, \lambda)$,
and we define the effective degree, a scalar important 
for spreading mechanisms, 
as $\keff (\nu, \lambda) = \kout(\nu, \lambda) +\kbid(\nu, \lambda)$.
We also introduce a directedness indicator function 
$d(\lambda)$
which equals one if edges of type $\lambda$ are directed 
and zero if not.

To characterize a random network with
arbitrary node-edge-node correlations,
we need to specify a number of interrelated probabilities,
and these must further satisfy
certain restrictions and detailed balance equations~\cite{boguna2005a}.
First, we have the node and edge distributions
$\Probof{\nu}$
and
$\Probof{\lambda}$.
Note that we immediately have the restriction 
$\Probof{\lambda} = \Probof{\bar{\lambda}}$.
Also, these induce the usual degree distributions via
\[
\Probof{\veck} = 
\sum_{\nu \in \mathcal{N}}
\Probof{\nu} \delta_{\veck, \veck(\nu) }
\]
where $\delta$ is the Kronecker delta.

Next we need 
$\Probof{\nu\lambda}$,
defined as the probability 
that, in randomly choosing an edge
and traversing it 
(in the allowed direction if directed or a random direction if undirected),
we find it is of type $\lambda$ and
that we are travelling away from a node of type $\nu$.

Finally, we encode correlations via the transition probability
$\Probof{\nu|\nu'\lambda'}$ which is the probability that
we reach a type $\nu$ node,
given that we are following a
type $\lambda'$ edge away
from a type $\nu'$ node.
This includes the usual degree-degree transition probabilities
(see Sec.~\ref{subsec:gcgrn.applicationrandom} and 
\cite{payne2011a} for notation):
\begin{equation}
  \label{eq:gcgrn.gc-grn.transitionprdeg}
\begin{aligned}
  \Probin(\veck''|\veck') &= 
  \sum_{\nu, \nu' \in \mathcal{N}}
  \sum_{\lambda \in \Lambda_{\nu'}}
  \Probof{\nu | \nu'\lambda'}
  d(\lambda')
  \\
  &\qquad\times
  \delta_{\veck',\veck(\nu')}
  \delta_{\veck'',\veck(\nu)},
  \\
  \Probbid(\veck''|\veck') &= 
  \sum_{\nu, \nu' \in \mathcal{N}}
  \sum_{\lambda \in \Lambda_{\nu'}}
  \Probof{\nu | \nu'\lambda'}
  [1-d(\lambda')]
  \\
  &\qquad\times
  \delta_{\veck',\veck(\nu')}
  \delta_{\veck'',\veck(\nu)}.
\end{aligned}
\end{equation}

We are now forced to connect and constrain the probabilities 
$\Probof{\nu\lambda}$
and
$\Probof{\nu|\nu'\lambda'}$
according to a detailed balance constraint.
Consider $\Probof{\nu'\lambda'\nu}$ defined as
the probability that a randomly selected edge
is of type $\lambda'$ and
runs from a type $\nu'$ node to a type $\nu$ node
(corresponding to the subnetwork in Fig.~\ref{fig:gcgrn.sketch}).
Then,
\[
\Probof{\nu'\lambda'\nu} = \Probof{\nu|\nu'\lambda'} \Probof{\nu' \lambda'}.
\]
Now, if we traversed the edge in the disallowed direction, 
it would ``connect'' a type $\nu$ node to a type $\nu'$ node.
Then we must also have
$\Probof{\nu'\lambda'\nu} = \Probof{\nu\bar{\lambda'}\nu'}$.
We therefore arrive at the detailed balance condition:
\begin{equation}
  \label{eq:gcgrn.gc-grn.detailedbalance}
  \underbrace{
    \Probof{\nu|\nu'\lambda'}
    \Probof{\nu'\lambda'}
  }_{\Probof{\nu'\lambda'\nu}}
  =
  \underbrace{
    \Probof{\nu'|\nu\bar{\lambda'}}
    \Probof{\nu\bar{\lambda'}}
  }_{\Probof{\nu\bar{\lambda'}\nu'}}.
\end{equation}

Note that the detailed balance condition, 
Eq.~\eqref{eq:gcgrn.gc-grn.detailedbalance},
is more general for typed random networks
than the detailed balance conditions
in terms of the degree distributions 
$\Probof{\veck,\veck'}$ and $\Probof{\veck}$
found by \cite{boguna2005a}.
If the types of the nodes are their degrees
and the edge types are 
$\Lambda = \{\rm undirected, incoming, outgoing\}$,
then Eq.~\eqref{eq:gcgrn.gc-grn.detailedbalance}
reduces to the well-known detailed balance conditions
given in \cite{boguna2005a} and \cite{payne2011a},
which can all be written as
\begin{equation}
  \label{eq:gcgrn.gc-grn.detailedbalanceks}
  \underbrace{
    P^{(\lambda)}(k | k') \frac{k_\lambda' \Probof{k'}}{\langle k_\lambda \rangle} 
  }_{P^{(\lambda)}(k,k')}
  =
  \underbrace{
    P^{(\bar{\lambda})}(k' | k) \frac{k_{\bar{\lambda}} \Probof{k}}{\langle k_\lambda \rangle}
  }_{ P^{(\bar{\lambda})}(k',k) } .
\end{equation}

In networks where there are multiple types of directed or undirected edges,
the detailed balance equations given in
\cite{boguna2005a, payne2011a},
which have the form of Eq.~\eqref{eq:gcgrn.gc-grn.detailedbalanceks},
are not necessarily valid.
This is because not all edges or degree-$k$ nodes
are equivalent.
Using Eq.~\eqref{eq:gcgrn.gc-grn.detailedbalance},
we can show that the symmetry of the degree distributions
is conserved,
$\Probof{k,k'} = \Probof{k',k}$.
\[
\Probof{k'',k'} =
\sum_{\nu,\nu' \in \mathcal{N}} 
\sum_{\lambda \in \Lambda_\nu}
\Probof{\nu'\lambda\nu} 
\delta_{k',k(\nu')}
\delta_{k'',k(\nu)}
\]

In considering contagion processes,
we recall the well-known typical macroscopic 
`bow-tie' form of
random networks with directed edges~\cite{newman2001b,broder2000a,boguna2005a},
given that a giant component is present.
As shown in Fig.~\ref{fig:gcgrn.sketch},
there are three giant components of functional importance:
(1) the giant strongly connected component, $\Omega_{\rm scc}$,
within which any pair of nodes can be connected via
a path of directed and/or undirected edges,
traversing the directed ones;
(2) the giant in-component $\Omega_{\rm in}$,
the set of all nodes from which paths lead to $\Omega_{\rm scc}$
(n.b., $\Omega_{\rm scc} \subset \Omega_{\rm in}$);
and
(3) the giant out-component $\Omega_{\rm out}$,
the set of all nodes which can be reached along directed
paths starting from a node in $\Omega_{\rm in}$
(n.b., $\Omega_{\rm scc} \subset \Omega_{\rm out}$).
By definition, we have that $\Omega_{\rm scc} = \Omega_{\rm in} \cap \Omega_{\rm out}$.
Any global spreading event must begin
from a seed in the giant in-component,
and can at most spread to the giant out-component $\Omega_{\rm out}$.

\section{Generalized contagion process}
\label{sec:gcgrn.contagion}

We consider contagion processes where the probability
of a node's infection may depend in any fashion on the current
states of its neighbors, potentially resembling 
phenomena ranging from the spread of infectious diseases 
to socially-transmitted behaviors~\cite{schelling1973a,granovetter1978a,murray2002a,watts2002a}.
Since we are interested in the probability of spreading,
we can capitalize on the fact that random networks
are locally pure branching structures.  We therefore
need to know only what the probability of infection
is for a type $\nu$ node given a single neighbor
of type $\nu'$ is infected, whose influence is 
felt along a type $\lambda'$ edge.
We write this probability as $B_{\nu' \lambda' \nu}$.
Time is removed from this quantity, as we need
to know only the probability of eventual infection.
Disease spreading models with recovery~\cite{murray2002a,dodds2011b}
are included, as are threshold models inspired by
social contagion~\cite{granovetter1978a,watts2002a}.

\section{Triggering probabilities}
\label{sec:gcgrn.analysis}

\begin{table*}[tbp]
  \centering
  \begin{tabular}{
      |>{\PBS\raggedright\hspace{0pt}}m{0.13\textwidth}
      |>{\PBS\raggedright\hspace{0pt}}m{0.5\textwidth}
      |>{\PBS\raggedright\hspace{0pt}}m{0.36\textwidth}
      |}
    \toprule
    \textbf{Network:} & 
    \textbf{Edge Triggering Probability:} & 
    \textbf{Node Triggering Probability, $Q$:} \\
    \colrule
    & & \\
    I. Undirected, Uncorrelated & 
    $
    \displaystyle
    \probfc_{\ast \ast} = 
    \sum_{\kbid}
    \Probbid(\kbid\, | \, \ast)
    B_{\ast \ast \kbid}
    \left[
      1
      -
      (1-\probfc_{\ast \ast})^{\kbid-1}
    \right]
    $
    &
    $
    \displaystyle
    \sum_{\kbid}
    \Probof{\kbid}
    \left[
      1 - (1-\probfc_{\ast \ast})^{\kbid}
    \right]
    $ 
    \\
    & & \\
    II. Directed, Uncorrelated & 
    $
    \displaystyle
    \probfc_{\ast \ast} = 
    \sum_{\kin,\kout}
    \Probbid(\kin,\kout | \, \ast)
    B_{\ast \ast \kin}
    \left[
      1
      -
      (1-\probfc_{\ast \ast})^{\kout}
    \right]
    $
    &
    $
    \displaystyle
    \sum_{\kin,\kout}
    \Probof{\kin,\kout}
    \left[
      1 - (1-\probfc_{\ast \ast})^{\kout}
    \right]
    $ \\
    & & \\
    III. Mixed Directed and Undirected, Uncorrelated & 
    $
    \displaystyle
    \probfc_{\ast u}
    = 
    \sum_{\veck}
    \Probbid(\veck | \, \ast)
    B_{\ast \ast \veck}
    \left[
      1
      -
      (1-\probfc_{\ast u})^{\kbid-1}
      (1-\probfc_{\ast o})^{\kout}
    \right]
    $
    \newline
    $
    \displaystyle
    \probfc_{\ast o} 
    = 
    \sum_{\veck}
    \Probin(\veck | \, \ast)
    B_{\ast \ast \veck}
    \left[
      1
      -
      (1-\probfc_{\ast u})^{\kbid}
      (1-\probfc_{\ast o})^{\kout}
    \right]
    $
    & 
    $
    \displaystyle
    \sum_{\veck}
    \Probof{\veck}
    \left[
      1 - (1-\probfc_{\ast u})^{\kbid}(1-\probfc_{\ast o})^{\kout}
    \right]
    $ 
    \\
    & & \\
    \colrule
    & & \\
    IV. Undirected, Correlated & 
    $
    \displaystyle
    \probfc_{\kbid' \ast} = 
    \sum_{\kbid}
    \Probbid(\kbid\, | \, \kbid')
    B_{\ast \ast \kbid}
    \left[
      1
      -
      (1-\probfc_{\kbid \ast})^{\kbid-1}
    \right]
    $
    &
    $
    \displaystyle
    \sum_{\kbid} 
    \Probof{\kbid}
    \left[
      1 - (1-\probfc_{\kbid \ast})^{\kbid}
    \right]
    $ 
    \\
    & & \\
    V. Directed, Correlated & 
    $
    \displaystyle
    \probfc_{\kin'\kout',\ast} = 
    \sum_{\kin,\kout}
    \Probbid(\kin,\kout | \, \kin',\kout')
    B_{\ast \ast \kin}
    \left[
      1
      -
      (1-\probfc_{\kin\kout,\ast})^{\kout}
    \right]
    $
    &
    $
    \displaystyle
    \sum_{\kin,\kout} 
    \Probof{\kin,\kout}
    \left[
      1 - (1-\probfc_{\kin\kout, \ast})^{\kout}
    \right]
    $ \\
    & & \\
    VI. Mixed Directed and Undirected, Correlated & 
    $
    \displaystyle
    \probfc_{\veck' u} 
    = 
    \sum_{\veck}
    \Probbid(\veck | \, \veck')
    B_{\ast \ast \veck}
    \left[
      1
      -
      (1-\probfc_{\veck u})^{\kbid-1}
      (1-\probfc_{\veck o})^{\kout}
    \right]
    $
    \newline
    $
    \displaystyle
    \probfc_{\veck' o} 
    = 
    \sum_{\veck}
    \Probin(\veck | \, \veck')
    B_{\ast \ast \veck}
    \left[
      1
      -
      (1-\probfc_{\veck u})^{\kbid}
      (1-\probfc_{\veck o})^{\kout}
    \right]
    $
    & 
    $
    \displaystyle
    \sum_{\veck} 
    \Probof{\veck}
    \left[
      1 - (1-\probfc_{\veck u})^{\kbid}
      (1-\probfc_{\veck o})^{\kout}
    \right]
    $ 
    \\
    & & \\
    \botrule
  \end{tabular}
  \caption{
    For the six classes of random networks described
    in Sec.~\ref{subsec:gcgrn.applicationrandom},
    the probability of triggering a global spreading events
    due to
    (1) an infected edge,
    and
    (2) an infected, randomly chosen single seed
    (see Eqs.~\ref{eq:gcgrn.edgetrig-general} and~\ref{eq:gcgrn.nodetrig-general}).
    We indicate by the symbol $\ast$ 
    when no node or edge type is relevant.
    }
  \label{tab:gcgrn.trigprobs}
\end{table*}

We define $Q_{\nu\lambda}$ to be the probability
that seeding a type $\nu$ node generates a global
spreading event along an edge of type $\lambda$.
Due to the Markovian nature of random networks,
this probability must satisfy a nonlinear
recursion relation: 
\begin{equation}
  \label{eq:gcgrn.edgetrig-general}
  \begin{aligned}
  Q_{\nu'\lambda'}
  &=
  \sum_{\nu \in \mathcal{N}}
  \Probof{\nu | \nu'\lambda'}
  B_{\nu' \lambda' \nu} \\
  &\;\;\;
  \times
  \left[
    1 
    -
    \prod_{ \lambda \in \Lambda_{\nu} }
    \left(
      1
      -
      Q_{\nu\lambda}
    \right)^{
      \keff(\nu,\lambda) - \delta_{\lambda,\bar{\lambda'}}
    }
  \right],
  \end{aligned}
\end{equation}
an expression which involves the following three elements.
First, we have $\Probof{\nu | \nu'\lambda'}$
which is the probability of transitioning to a node of type $\nu$.
The second term $B_{\nu'\lambda'\nu}$
is the probability of successful infection.  
The last term contains the recursive structure.
At least one of the edges leading away from the
type $\nu$ node must generate a global spreading
event 
(note that we avoid double counting
the incident edge of type $\bar{\lambda'}$
with the indicator in the exponent).  
The probability this happens is 
the complement of the probability that none succeed,
$
\prod_{
  \lambda \in \Lambda_{\nu}
}
\left(
  1
  -
  Q_{\nu \lambda }
\right)^{
      \keff(\nu,\lambda) - \delta_{\lambda,\bar{\lambda'}}
    }.
$
\Req{eq:gcgrn.edgetrig-general} will rarely be 
analytically tractable 
(but see~\cite{payne2011a} for
an exactly solved simple model), 
and will usually be 
solved numerically by iteration.

The probability that an infected type $\nu$ node seeds
a global spreading event follows as
\begin{equation}
  \label{eq:gcgrn.nodetrig-specific}
  Q_{\nu}  
  =
  1-
  \prod_{\lambda \in \Lambda_{\nu}}
  (1 - Q_{\nu\lambda})^{
    \keff(\nu,\lambda) 
    },
\end{equation}
where again success is defined in terms of not failing.
Finally, the probability that the sole infection of
a randomly chosen node leads to a global spreading event is
\begin{equation}
  \begin{aligned}
  Q
  =
  \sum_{\nu' \in \mathcal{N}}
  \Probof{\nu'}
  Q_{\nu'}                       \end{aligned}
  \label{eq:gcgrn.nodetrig-general}
\end{equation}
The effects of weighted triggering schemes---where the initial
node is chosen according to its degree in some fashion---can be
easily examined by replacing $\Probof{\nu'}$ with the appropriate distribution.

\section{Connection between triggering probabilities and the contagion condition}
\label{sec:gcgrn.contagioncondition}

We show how our general expression for triggering probabilities
reduces to the general cascade condition we described in~\cite{dodds2011b}.
The calculation involved makes an important analytic connection
but is necessarily largely mathematical in nature, 
as represented in Fig.~\ref{fig:gcgrn.sketch}.

The cascade condition is a binary expression of possibility; when
the condition is met, global spreading events initiated by single seeds 
are possible, and otherwise they are impossible.
Starting from~\Req{eq:gcgrn.edgetrig-general}, 
we  determine the cascade condition by examining under
what circumstances the triggering probability 
$Q_{\nu\lambda} \rightarrow 0^{+}$.
In this limit, the product in~\Req{eq:gcgrn.edgetrig-general},
$
\prod_{
  \lambda \in \Lambda_{\nu}
}
\left(
  1
  -
  Q_{\nu \lambda }
\right)^{
      \keff(\nu,\lambda) - \delta_{\lambda,\bar{\lambda'}}
    }.
$
can be approximated as 
$
1 - 
\sum_{
  \lambda \in \Lambda_{\nu}
}
\left(
  \keff(\nu,\lambda) - \delta_{\lambda,\bar{\lambda'}}
\right)
Q_{\nu\lambda},
$
to first order.
Neglecting higher order terms,
\Req{eq:gcgrn.edgetrig-general} reduces to
\begin{equation}
  \label{eq:gcgrn.edgetrig-general-approx}
  \begin{aligned}
  Q_{\nu'\lambda'}
  &\simeq
  \sum_{\nu \in \mathcal{N}}
  \Probof{\nu | \nu'\lambda'}
  B_{\nu |\lambda'\nu'} \\
  & \quad \times \sum_{
    \lambda \in \Lambda_{\nu}
  }
  \left(
    \keff(\nu,\lambda) - \delta_{\lambda,\bar{\lambda'}}
  \right)
  Q_{\nu \lambda}.
  \end{aligned}
\end{equation}
We introduce the notation from~\cite{dodds2011b},
where
$\alpha=(\nu,\lambda)$
and
$\alpha'=(\nu',\lambda')$,
as well as 
$k_{\alpha' \alpha} = 
    \keff(\nu,\lambda) - \delta_{\lambda,\bar{\lambda'}}
$
as the number of type $\lambda$ edges
leaving from nodes of type $\nu$,
with the exclusion of the incident type $\lambda'$ 
edge arriving from a type $\nu'$ node.
We also let 
$P_{\alpha' \alpha} =
\Probof{\nu|\nu' \lambda'}$,
$B_{\alpha' \alpha} = B_{\nu'\lambda'\nu}$.
Note that the outgoing edge of type $\lambda'$ does
not affect the contagion mechanism and is left as arbitrary
in $\alpha'$.
Then the above equation becomes
\begin{equation}
  \label{eq:gcgrn.edgetrig-general-approx-alpha}
  Q_{\alpha'}
  \simeq
  \sum_{\alpha}
  P_{\alpha' \alpha}
  \bullet
  k_{\alpha' \alpha}
  \bullet
  B_{\alpha' \alpha}
  Q_{\alpha}
  = 
    \sum_{\alpha}
  R_{\alpha' \alpha}
  Q_{\alpha},
\end{equation}
where we have identified the  gain matrix $R$
we obtained and described in~\cite{dodds2011b}.
Contagion is possible only when the largest eigenvalue
of $R$ exceeds unity,
and we have connected the triggering probability to the cascade condition.

\section{Applications}

\subsection{Triggering probabilities for six random network families}
\label{subsec:gcgrn.applicationrandom}

In Tab.~\ref{tab:gcgrn.trigprobs},
we list the forms of $Q_{\nu' \lambda'}$ and $Q$ for
six specific families of random networks
which we describe below.
The last of these network families
is the most general and contains the other five
as special cases.  
Nodes potentially have three
kinds of unweighted edges incident to them: undirected, in-directed,
and out-directed, and we use the vector representation
$\vec{k} = \left( \kbid,\kin,\kout \right)$
to define node classes~\cite{boguna2005a,dodds2011b}.
The specific transition probabilities, 
$\Probin(\veck|\veck')$,
$\Probout(\veck|\veck')$,
and
$\Probbid(\veck|\veck')$,
give the probabilities of an edge leading from
a degree $\veck'$ node to a degree $\veck$ node being
oriented as undirected, incoming, or outgoing
(see Refs.~\cite{dodds2011b} and~\cite{payne2011a}
for more details).
For uncorrelated networks, we use
the notation $\Probin(\veck|\,\ast)$, etc.
Similarly for the triggering probabilities,
where the node or edge type is irrelevant
we also use $\ast$ (e.g., $Q_{\ast \ast}$ instead
of $Q_{\nu \lambda}$ for undirected, uncorrelated, unweighted networks).
For simplicity, we assume infection is due only to properties
of the node potentially being infected, which for these networks means 
the node's degree.

\subsection{Random bipartite networks}
\label{subsec:gcgrn.applicationbipartite}

We now show how the theory of contagion in 
bipartite networks~\cite{newman2001b}
is a special case of the general model.
Consider a bipartite network
$G = (V,E)$
with the nodes partitioned
into disjoint sets
$\setone$
and
$\settwo$,
such that 
$V = \setone \cup \settwo$
and all edges 
$uv \in E$
satisfy 
$u \in \setone$ and $v \in \settwo$
or
$u \in \settwo$ and $v \in \setone$.
Again, we consider general node types
$\nodetype$,
but now they are also associated with
either one of the sets
$\setone$ or $\settwo$.

Due to the bipartite structure, 
the triggering probability~\Req{eq:gcgrn.edgetrig-general}
separates into two coupled equations
\begin{equation}
  \label{eq:gcgrn.bipartitenonlin}
\begin{aligned}
  Q_{\nodetype' \edgetype'}^\supone &= 
  \sum_{\nodetype} \Prob^\supone( \nodetype| \nodetype' \edgetype')
  B^\supone_{ \nodetype' \edgetype' \nodetype }
  \\
  & \times
  \left[ 
    1 - 
    \prod_{
      \lambda \in \Lambda_{\nu}
    }
    \left( 
      1 - Q_{\nodetype \edgetype}^\suptwo 
    \right)^{
      \keff(\nu,\lambda) - \delta_{\lambda,\bar{\lambda'}}
    }
  \right], 
  \\
  Q_{\nodetype' \edgetype'}^\suptwo &=
  \sum_{\nodetype} \Prob^\suptwo( \nodetype | \nodetype' \edgetype')
  B^\suptwo_{ \nodetype' \edgetype' \nodetype }
  \\
  & \times
  \left[ 
    1 - 
    \prod_{
      \lambda \in \Lambda_{\nu}
    }
    \left( 
      1 - Q_{\nodetype \edgetype}^\supone 
    \right)^{
      \keff(\nu,\lambda) - \delta_{\lambda,\bar{\lambda'}}
    }
  \right] ,
\end{aligned}
\end{equation}
where the superscripts denote the triggering probabilities 
starting in
$\setone$ and $\settwo$, respectively.

The contagion condition arises again by linearizing
Eq.~\eqref{eq:gcgrn.bipartitenonlin} about 
$Q^\supone = Q^\suptwo = 0$.
This gives the linear system of equations
\begin{equation}
  \label{eq:gcgrn.bipartitelin1}
  \begin{aligned}
    Q_{\nodetype' \edgetype'}^\supone &=
    \sum_{\nodetype} 
    \sum_{ \edgetype \in \Lambda_{\nodetype} }
    \Prob^\supone(\nodetype|  \nodetype' \edgetype')
    B^\supone_{ \nodetype' \edgetype' \nodetype } \\
    & \times
    \left(
      \keff(\nu,\lambda) - \delta_{\lambda,\bar{\lambda'}}
    \right)
    Q_{\nodetype \edgetype}^\suptwo,  
  \end{aligned}
\end{equation}
\begin{equation}
  \label{eq:gcgrn.bipartitelin2}
  \begin{aligned}
    Q_{\nodetype' \edgetype'}^\suptwo &=
    \sum_{\nodetype} 
    \sum_{ \edgetype \in \Lambda_{\nodetype} }
    \Prob^\suptwo(\nodetype|  \nodetype' \edgetype')
    B^\suptwo_{ \nodetype' \edgetype' \nodetype }\\
    &\times
    \left(
      \keff(\nu,\lambda) - \delta_{\lambda,\bar{\lambda'}}
    \right)
    Q_{\nodetype \edgetype}^\supone.
  \end{aligned}
\end{equation}
These equations are of the form
\begin{equation}
  \left[
    \begin{array}{c}
      Q^\supone \\
      Q^\suptwo
    \end{array}
  \right]
  =
  \left[
    \begin{array}{cc}
      0 & R_{12} \\
      R_{21} & 0 \\
    \end{array}
  \right]
  \left[
    \begin{array}{c}
      Q^\supone \\
      Q^\suptwo
    \end{array}
  \right[
   = 
   R
  \left[
    \begin{array}{c}
      Q^\supone \\
      Q^\suptwo
    \end{array}
  \right],
\end{equation}
where the entries of $R_{12}$ and $R_{21}$ are shown in 
Eqs.~\eqref{eq:gcgrn.bipartitelin1} and~\eqref{eq:gcgrn.bipartitelin2}.
The structure of the gain matrix $R$, 
of course, reflects the bipartiteness of $G$.
Spreading will occur when the spectral radius $\rho(R) > 1$~\cite{dodds2011b}.
The eigenvalues of $R$ are the solutions $\lambda$ to
\[
\det ( \lambda^2 I - R_{12} R_{21} ) = 0,
\]
since the diagonal matrix 
$\lambda I$ 
and
$R_{21}$
commute~\cite{silvester2000a}.
The eigenvalues of $R$ are thus the square roots
of the eigenvalues of $R_{12} R_{21}$, 
meaning we can also express the contagion condition as
$\rho( R_{12} R_{21} ) > 1$.

There is a physical explanation for the contagion condition.
Assume the contagion starts with one active node in 
$\setone$.
It then must pass to 
$\settwo$
before returning to 
$\setone$.
The gain going from
$\setone$ to $\settwo$
is
$R_{12}$,
and the gain is
$R_{21}$
going from
$\settwo$ to
$\setone$.
If the expected number of active nodes after these two passes exceeds
unity, the contagion will spread.
Note that the spectra of $R_{12} R_{21}$ and $R_{21} R_{12}$
are equal, so that we could also consider starting the contagion
in $\settwo$.

\subsection{Uncorrelated, undirected bipartite networks}
\label{subsec:gcgrn.simplebipartite}

We now confirm that the general theory gives the previously 
known results for  uncorrelated, undirected bipartite networks.
These networks are fully specified by the degree distributions
$P^\supone (k)$ and $P^\suptwo (k)$
for nodes in sets
$\setone$ and $\settwo$,
respectively.
We set the infection probability $B^\supone = B^\suptwo = 1$ for all nodes,
so that we are solving for the existence of a giant component.
The edge probabilities are
\begin{align}
  P^\supone (k | *) &= 
  \frac{k P^\supone (k)}{\sum_{k} k P^\supone (k) }
  \label{eq:gcgrn.bipartiteCondProb1} \\
  P^\suptwo (k | *) &= 
  \frac{k P^\suptwo (k)}{\sum_{k} k P^\suptwo (k) }
  \label{eq:gcgrn.bipartiteCondProb2}
\end{align}
where 
$P^\supone(k | *)$ 
is the probability of reaching a degree $k$
node in $\setone$ from a random node in $\settwo$, and 
$P^\suptwo(k' | *)$ 
is likewise the probability of reaching a degree $k'$ 
node in $\settwo$ from a random
node in $\setone$.

Pick a random node 
$u \in \setone$
and imagine that the
contagion arrives at $u$
via one of its incoming edges.
Then there are an expected
$\sum_{k} (k-1) P^\supone(k | *) = R_{12}$
edges leftover, each leading to an unexplored node in $\settwo$.
Follow one of these to 
$ v \in \settwo$,
then the expected excess edges coming from
$v$
is
$\sum_{k'} (k'-1) P^\suptwo(k' | *) = R_{21}$.
Multiplying these two sums together gives the
expected number of new nodes reached in $\setone$
after two passes, so the contagion condition is
\begin{equation}
  R_{12} R_{21} 
  = 
  \sum_{k,k'} (k-1) P^\supone(k|*) (k'-1) P^\suptwo(k'|*) > 1 .
  \label{eq:gcgrn.bipartitecondition}
\end{equation}
Substituting \eqref{eq:gcgrn.bipartiteCondProb1} and 
\eqref{eq:gcgrn.bipartiteCondProb2} for the conditional
probabilities, taking the normalization factors
to the right hand side, and simplifying, we arrive at
\begin{equation}
  \sum_{k,k'} k k' (k k'-k-k') P^\supone(k) P^\suptwo(k') > 0
  \label{eq:gcgrn.bipartiteNewman}
\end{equation}
which is the condition found by Newman, Strogatz, and Watts
\cite{newman2001b} using generating functions.
While 
Eqs.~\ref{eq:gcgrn.bipartitecondition}
and~\ref{eq:gcgrn.bipartiteNewman}
are equivalent, the former preserves
the physics of the spreading process.

\section{Concluding remarks}
\label{sec:gcgrn.conclusion}

We have shown that the probability of
a single infected node generating a global
spreading event can be derived in a straightforward way
for spreading processes on a very general class of 
correlated random networks.
Our approach brings a physical intuition to 
the problem, and while more sophisticated mathematical
analyses arrive at the same results, and are certainly
useful for more detailed investigations, they are
burdened with some degree of inscrutability.

\acknowledgments
We appreciate discussions with Braden Brinkman.
KDH was supported by VT-NASA EPSCoR and a Boeing fellowship;
JLP was supported by NIH grant \# K25-CA134286;
PSD was supported by NSF CAREER Award \# 0846668.

\appendix

\end{document}